\documentclass{article}
\usepackage[latin9]{inputenc}
\usepackage{geometry}
\usepackage{textcomp}
\usepackage{amsmath}
\usepackage{amssymb}
\usepackage[unicode=true,pdfusetitle,
 bookmarks=true,bookmarksnumbered=false,bookmarksopen=false,
 breaklinks=false,pdfborder={0 0 1},backref=section,colorlinks=false]
 {hyperref}
\usepackage{breakurl}

\makeatletter

\providecommand{\tabularnewline}{\\}

\usepackage{float}
\usepackage{graphicx}
\usepackage{grffile}
\usepackage{breakurl}

\providecommand{\tabularnewline}{\\}

\usepackage{fullpage}\usepackage[justification=centering]{caption}\usepackage[all]{hypcap}\usepackage{microtype}

\makeatother

\begin{document}

\title{J=0 T=1 Pairing Interaction Selection Rules}

\author{Matthew Harper and Larry Zamick\\
 \textit{Department of Physics and Astronomy, Rutgers University,
Piscataway, New Jersey 08854}\\
 }

\date{\today}

\maketitle
 
\begin{abstract}
Wave functions arising form a pairing Hamiltonian E(0) i.e. one in
which the interaction is only between J=0$^{+}$ T=1 pairs, lead to
magnetic dipole and Gamow-Teller transition rates that are much larger
than those from an interaction E(J$_{\text{max}}$) in which a proton
and a neutron couple to J=2$j$. With realistic interactions the results
are in between the 2 extremes. In the course of this study we found
that certain M1 and GT matrix elements vanish with E(0). These are
connected to seniority and reduced isospin selection rules. We find
the surprising result that the M1 strength to the ``single j scissors''
is larger for a J=0 T=1 pairing interaction than it is for Q.Q. 
\end{abstract}

\section{Introduction}

We have recently performed single $j$ shell studies of both schematic
and realistic interactions{[}1{]}. They ranged from the J=0$^{+}$
T=1 to the J$_{\text{max}}$ T=0 interactions. In this work we will
focus more on the experimental consequences of choosing a given interaction.
In particular we study Gamow-Teller and isovector M1 matrix elements
for transitions in Sc and Ti isotopes. Some of the problems have been
addressed numerically in a previous publication {[}2{]}, but here
we will present analytical proofs.

\subsection{The Interactions}

For two particles in a single j shell the states of even angular momentum
J have isospin T=1 and those of odd J have T=0. For convenience; we
define E(J) as a two body interaction which is zero except when the
two particles couple to J. Hence, we have the J=0$^{+}$ T=1 pairing
interaction designated as E(0) and the other extreme E(J$_{\text{max}}$)
which acts only in the T=0 state with J$_{\text{max}}$ =2j. The T=0
odd J interaction acts only between a neutron and a proton. We only
consider charge independent interactions in this work. For a ``realistic''
interaction in the f$_{7/2}$ shell we use the MBZE interaction {[}3{]}.
This is based on the works of Bayman et al. and McCullen et al. {[}4,5{]}
but with improved T=0 two-body matrix elements{[}4{]}. From J=0 to
J$_{\text{max}}$=7 the matrix elements, which were obtained from
experiment are: 
\begin{gather}
(0.0000,0.6111,1.5863,1.4904,2.8153,1.5101,3.2420,0.6163)
\end{gather}
Although the J=0$^{+}$ matrix element is the most attractive; in
MBZE one also has low lying T=0 levels with J=1$^{+}$ and J=J$_{\text{max}}=7^{+}$.
Indeed, one main thrust of the old papers was that there was a large
probability in say, an even-even nucleus that the protons and neutrons
do not couple to zero. Indeed; it was shown in ref {[}5{]} that a
much better overlap with the realistic interaction was obtained with
a quadrapole-quadrapole interaction(Q.Q) than with the J=0 pairing
interaction. We should also mention here the work on GT by Lawson{[}6{]}
who invoked a K selection rule to explain why GT matrix elements decrease
with neutron excess.

\section{Wave Functions And Quantum Numbers for a J=0 T=1 Pairing Interaction
of a Q.Q. Interaction}

In this section we present energy levels and wave functions of $^{43}$Sc
and $^{44}$Ti which have a J=0 T=1 pairing interaction of Flowers
and Edmonds {[}7,8{]} and a Q.Q. interaction. The wave functions are
presented as column vectors of probability amplitudes. To identify
the higher isospin states we subtracted 3 MeV from all T=0, two-body
matrix elements for the pairing interaction. Doing so does not affect
the wave functions of the non-degenerate states, but it will remove
degeneracies of states with different isospins. For Scandium isotopes
we indicate a star $(*)$ for states with T=3/2. For $^{44}$Ti we
indicate a star for T=1 and two stars $(**)$ for T=2.

\subsection*{Table 2.1 Energies(MeV) and Wave Functions of $^{43}$Sc with a J=0
T=1 Pairing Interaction}

I=5/2

\centering 
\begin{tabular}{rrrrr}
\hline 
J$_{p}$  & J$_{n}$  & 1.125  & 1.125  & 5.625$^{*}$\tabularnewline
\hline 
3.5  & 2.0  & 0.4210  & -0.4600  & 0.7817\tabularnewline
3.5  & 4.0  & 0.4695  & 0.8479  & 0.2462\tabularnewline
3.5  & 6.0  & 0.7761  & -0.2633  & -0.5730 \tabularnewline
\hline 
\end{tabular}\\

\raggedright I=7/2

\centering 
\begin{tabular}{rrrrrr}
\hline 
J$_{p}$  & J$_{n}$  & 0.000  & 1.125  & 1.125  & 4.875$^{*}$\tabularnewline
\hline 
3.5  & 0.0  & 0.8660  & 0.000  & 0.000  & 0.500\tabularnewline
3.5  & 2.0  & 0.2152  & -0.8924  & -0.1358  & 0.3727\tabularnewline
3.5  & 4.0  & 0.2887  & 0.1565  & 0.8014  & 0.500\tabularnewline
3.5  & 6.0  & 0.3469  & 0.4232  & -0.5826  & 0.6009 \tabularnewline
\hline 
\end{tabular}\\

\raggedright I=9/2

\centering %
\begin{tabular}{rrrrr}
\hline 
J$_{p}$  & J$_{n}$  & 1.125  & 1.125  & 5.625$^{*}$ \tabularnewline
\hline 
3.5  & 2.0  & -0.1015  & 0.9416  & -0.3212\tabularnewline
3.5  & 4.0  & 0.4930  & 0.3280  & .08058\tabularnewline
3.5  & 6.0  & 0.8641  & -0.0766  & -0.4975\tabularnewline
\hline 
\end{tabular}\\
 \raggedright

\subsection*{Table 2.2 Energies(MeV) and Wave Functions of $^{44}$Ti with a J=0
T=1 Pairing Interaction}

I=0

\centering 
\begin{tabular}{rrrrrr}
\hline 
J$_{p}$  & J$_{n}$  & 0.000  & 0.750$^{**}$  & 2.25  & 2.25 \tabularnewline
\hline 
0.0  & 0.0  & 0.8660  & -0.5000  & 0.000  & 0.000\tabularnewline
2.0  & 2.0  & 0.2152  & 0.3737  & 0.8863  & 0.1712\tabularnewline
4.0  & 4.0  & 0.2887  & 0.5000  & -0.1244  & -0.8070\tabularnewline
6.0  & 6.0  & 0.3469  & 0.6009  & -0.4461  & 0.5652\tabularnewline
\hline 
\end{tabular}\\

\raggedright I=1

\centering %
\begin{tabular}{rrrrr}
\hline 
J$_{p}$  & J$_{n}$  & 1.500$^{*}$  & 2.250$^{*}$  & 2.250$^{*}$ \tabularnewline
\hline 
2.0  & 2.0  & 0.1992  & 0.9258  & 0.3212\tabularnewline
4.0  & 4.0  & 0.4879  & -0.3780  & 0.7868\tabularnewline
6.0  & 6.0  & 0.8498  & 0.0000  & -0.5270\tabularnewline
\hline 
\end{tabular}\\

\raggedright \newpage{}

\raggedright I=2

\centering %
\begin{tabular}{rrrrrrrrrrr}
\hline 
J$_{p}$  & J$_{n}$  & 1.000  & 1.250  & 1.750  & 2.250  & 2.250  & 2.250  & 2.250  & 2.250  & 2.250 \tabularnewline
\hline 
0.0  & 2.0  & 0.6455  & 0.7071  & -0.2887  & 0.0000  & 0.0000  & 0.0000  & 0.0000  & 0.0000  & 0.0000\tabularnewline
2.0  & 0.0  & 0.6455  & -0.7071  & -0.2887  & 0.0000  & 0.0000  & 0.0000  & 0.0000  & 0.0000  & 0.0000\tabularnewline
2.0  & 2.0  & -0.1205  & 0.0000  & -0.2694  & 0.6032  & 0.3665  & -0.0549  & -0.0618  & -0.3799  & 0.5134\tabularnewline
2.0  & 4.0  & 0.1730  & 0.0000  & 0.3869  & -0.1407  & -0.4053  & -0.0033  & 0.2281  & -0.7442  & 0.1746\tabularnewline
4.0  & 2.0  & 0.1730  & 0.0000  & 0.3869  & 0.6458  & 0.1122  & 0.1532  & 0.1480  & -0.0348  & -0.5867\tabularnewline
4.0  & 4.0  & -0.0193  & 0.0000  & -0.0431  & 0.0193  & 0.0946  & -0.5433  & 0.8105  & 0.1821  & 0.0569\tabularnewline
4.0  & 6.0  & 0.1403  & 0.0000  & 0.3138  & 0.3245  & -0.4415  & -0.5108  & -0.3715  & 0.3276  & 0.2746\tabularnewline
6.0  & 4.0  & 0.1403  & 0.0000  & 0.3138  & 0.0626  & -0.0068  & 0.5991  & 0.2948  & 0.3981  & 0.5230\tabularnewline
6.0  & 6.0  & 0.2292  & 0.0000  & 0.5125  & -0.2997  & 0.6964  & -0.2418  & -0.2013  & -0.0407  & 0.0973\tabularnewline
\hline 
\end{tabular}\\
 \raggedright I=2 (with shift in energy to remove degeneracies)

\centering %
\begin{tabular}{rrrrrrrrrrr}
\hline 
J$_{p}$  & J$_{n}$  & 1.000  & 2.250  & 2.250  & 2.250  & 4.250$^{*}$  & 5.250$^{*}$  & 5.250$^{*}$  & 10.750$^{**}$  & 11.250$^{**}$\tabularnewline
\hline 
0.0  & 2.0  & 0.6455  & 0.0000  & 0.0000  & 0.0000  & 0.7071  & 0.0000  & 0.0000  & -0.2887  & 0.0000\tabularnewline
2.0  & 0.0  & 0.6455  & 0.0000  & 0.0000  & 0.0000  & -0.7071  & 0.0000  & 0.0000  & -0.2887  & 0.0000\tabularnewline
2.0  & 2.0  & -0.1205  & 0.1561  & 0.6065  & 0.6391  & 0.0000  & 0.0000  & 0.0000  & -0.2694  & -0.3350\tabularnewline
2.0  & 4.0  & 0.1730  & -0.3895  & -0.1445  & 0.3056  & 0.0000  & -0.6977  & 0.1151  & 0.3869  & -0.2333\tabularnewline
4.0  & 2.0  & 0.1730  & -0.3895  & -0.1445  & 0.3056  & 0.0000  & 0.6977  & -0.1151  & 0.0869  & -0.2333\tabularnewline
4.0  & 4.0  & -0.0193  & 0.1797  & -0.3647  & 0.6726  & 0.0000  & 0.0000  & 0.0000  & -0.0431  & 0.6623\tabularnewline
4.0  & 6.0  & 0.1403  & -0.0861  & 0.4752  & -0.0728  & 0.0000  & 0.1151  & 0.6977  & 0.3138  & 0.3785\tabularnewline
6.0  & 4.0  & 0.1403  & -0.0861  & 0.4752  & -0.0728  & 0.0000  & 0.1151  & 0.6977  & 0.3138  & 0.3785\tabularnewline
6.0  & 6.0  & 0.2292  & 0.7906  & -0.0757  & 0.0165  & 0.0000  & 0.0000  & 0.0000  & 0.5125  & -0.2318\tabularnewline
\hline 
\end{tabular}\\
 \raggedright

\subsection*{Table 2.3 Energies(MeV) and Wave Functions of $^{43}$Sc with a Q.Q.
Interaction}

\raggedright I=5/2

\centering %
\begin{tabular}{rrrrr}
\hline 
J$_{p}$  & J$_{n}$  & 2.8243  & 3.0148  & 5.1306$^{*}$ \tabularnewline
\hline 
3.5  & 2.0  & 0.5053  & 0.7817  & -0.3655\tabularnewline
3.5  & 4.0  & 0.2885  & 0.2462  & 0.9253\tabularnewline
3.5  & 6.0  & 0.8133  & -0.5730  & -0.1011\tabularnewline
\hline 
\end{tabular}\\

\raggedright

I=7/2

\centering %
\begin{tabular}{rrrrrr}
\hline 
J$_{p}$  & J$_{n}$  & 0.000  & 3.3016$^{*}$  & 3.7874  & 5.4618 \tabularnewline
\hline 
3.5  & 0.0  & 0.7069  & -0.5000  & 0.4402  & 0.2376\tabularnewline
3.5  & 2.0  & 0.6864  & 0.3727  & -0.4393  & -0.4439\tabularnewline
3.5  & 4.0  & 0.1694  & 0.5000  & -0.1549  & 0.8350\tabularnewline
3.5  & 6.0  & 0.0216  & 0.6009  & 0.7676  & -0.2218\tabularnewline
\hline 
\end{tabular}\\
 \raggedright

I=9/2

\centering %
\begin{tabular}{rrrrr}
\hline 
J$_{p}$  & J$_{n}$  & 1.4765  & 4.1843  & 5.6367$^{*}$ \tabularnewline
\hline 
3.5  & 2.0  & 0.9032  & -0.2847  & -0.3212\tabularnewline
3.5  & 4.0  & 0.4186  & 0.4188  & 0.8058\tabularnewline
3.5  & 6.0  & 0.0949  & 0.8623  & -0.4975\tabularnewline
\hline 
\end{tabular}\\

\raggedright

\subsection*{Table 2.4 Energies(MeV) and Wave Functions of $^{44}$Ti with a Q.Q.
Interaction}

I=0

\centering %
\begin{tabular}{rrrrrr}
\hline 
J$_{p}$  & J$_{n}$  & 0.000  & 6.6031$^{**}$  & 7.5748  & 10.9236 \tabularnewline
\hline 
0.0  & 0.0  & 0.7069  & -0.5000  & 0.4402  & 0.2376\tabularnewline
2.0  & 2.0  & 0.6864  & 0.3727  & -0.4393  & -0.4439\tabularnewline
4.0  & 4.0  & 0.1694  & 0.5000  & -0.1549  & 0.8350\tabularnewline
6.0  & 6.0  & 0.0216  & 0.6009  & 0.7676  & -0.2218\tabularnewline
\hline 
\end{tabular}\\

\raggedright \newpage{}I=1

\centering %
\begin{tabular}{rrrrr}
\hline 
J$_{p}$  & J$_{n}$  & 4.3648 $^{*}$  & 7.3405 $^{*}$  & 10.5620$^{*}$\tabularnewline
\hline 
2.0  & 2.0  & 0.9109  & -0.2082  & -0.3563\tabularnewline
4.0  & 4.0  & 0.3967  & 0.2040  & 0.8950\tabularnewline
6.0  & 6.0  & 0.1137  & 0.9566  & -0.2684\tabularnewline
\hline 
\end{tabular}\\
 \raggedright I=2

\centering %
\begin{tabular}{rrrrrrrrrrr}
\hline 
J$_{p}$  & J$_{n}$  & 0.9665  & 4.6015$^{*}$  & 6.4691  & 7.7501  & 7.7502$^{**}$  & 8.5695$^{*}$  & 10.4893  & 10.6179$^{**}$  & 10.7351$^{*}$ \tabularnewline
\hline 
0.0  & 2.0  & 0.5807  & -0.5255  & 0.2263  & 0.08223  & -0.2887  & -0.4146  & 0.1466  & 0.0000  & 0.2280\tabularnewline
2.0  & 0.0  & 0.5807  & 0.5255  & 0.2263  & 0.08223  & -0.2887  & 0.4146  & 0.1466  & 0.0000  & -0.2280\tabularnewline
2.0  & 2.0  & -0.4331  & 0.0000  & 0.7001  & -0.2689  & -0.2694  & 0.0000  & 0.2554  & -0.3350  & 0.0000\tabularnewline
2.0  & 4.0  & 0.2513  & -0.4562  & 0.1629  & -0.35010  & 0.3869  & 0.3535  & -0.2881  & -0.2333  & 0.4085\tabularnewline
4.0  & 2.0  & 0.2513  & 0.4562  & 0.1629  & -0.35010  & 0.3869  & -0.3535  & -0.2880  & -0.2333  & 0.4805\tabularnewline
4.0  & 4.0  & -0.0916  & 0.0000  & 0.4892  & 0.1211  & -0.0431  & 0.0000  & -0.5451  & 0.6623  & 0.0000\tabularnewline
4.0  & 6.0  & 0.0403  & -0.1255  & 0.0802  & -0.2115  & 0.3138  & 0.4507  & 0.4533  & 0.3785  & 0.5302\tabularnewline
6.0  & 4.0  & 0.0403  & 0.1255  & 0.0802  & -0.2115  & 0.3138  & -0.4507  & 0.4533  & 0.3785  & -0.5302\tabularnewline
6.0  & 6.0  & -0.0099  & 0.0000  & 0.3198  & 0.75087  & 0.5125  & 0.0000  & 0.1334  & -0.2318  & 0.0000\tabularnewline
\hline 
\end{tabular}\\

\raggedright

\section{Assigning Quantum Numbers For J=0 T=1 Pairing}

e of the fact that we have the energies and wave functions of the
J=0$^{+}$ and J=1$^{+}$ states forming an explicit matrix digitalization.
It is convenient to add a constant so that the states which are not
collective are at zero energy. When this is done, the energies of
the J=0$^{+}$ $^{44}$Ti states are: 
\begin{gather}
-2.25,-1.5,0,0\text{ MeV}
\end{gather}
and the energies of the 1$^{+}$ states are: 
\begin{gather*}
-.75,0,0\text{ MeV}
\end{gather*}

We then fit these with the Flower's and Edmond's formula {[}7,8{]},
as given in Talmi's book{[}9{]}: 
\begin{gather}
\text{E}=C\left\lbrace \left(\frac{n-v}{4}\right)(4j+8-n-v)-\text{T}(\text{T}+1)+\text{t}(\text{t}+1)\right\rbrace 
\end{gather}
$C$ is most easily determined by the isospin splitting of the T=2
state at $-$1.5 MeV relative to the $-$2.25 ground state (in the
shifted energies). We set $-0.75=2\cdot3C$, so that $C=-0.125$ (In
general, $C=\frac{-1}{2j+1}$). The quantum numbers are shown in Tables
3.1 and 3.2. Previously; Neergaard {[}10{]} used this method to obtain
quantum numbers in his study of N=Z nuclei.\\

\subsection*{Table 3.1 Quantum Numbers for $^{43}$Sc with a Pairing Interaction}

\centering 
\begin{tabular}{rrrrr}
\hline 
Energy  & J  & T  & t  & v\tabularnewline
\hline 
0  & 5/2  & 1/2  & 1/2  & 3\tabularnewline
0  & 5/2  & 1/2  & 1/2  & 3\tabularnewline
0  & 5/2  & 3/2  & 3/2  & 3\tabularnewline
-1.125  & 7/2  & 1/2  & 1/2  & 1\tabularnewline
-0.75  & 7/2  & 3/2  & 1/2  & 1\tabularnewline
0  & 7/2  & 1/2  & 1/2  & 3\tabularnewline
0  & 7/2  & 1/2  & 1/2  & 3\tabularnewline
0  & 9/2  & 1/2  & 1/2  & 3\tabularnewline
0  & 9/2  & 1/2  & 1/2  & 3\tabularnewline
0  & 9/2  & 3/2  & 3/2  & 3\tabularnewline
\hline 
\end{tabular}\\
 \raggedright

\subsection*{Table 3.2 Quantum Numbers for $^{44}$Ti with a Pairing Interaction}

\centering J=0 %
\begin{tabular}{rrrr}
Energy  & T  & t  & v\tabularnewline
\hline 
-2.25  & 0  & 0  & 0\tabularnewline
-1.5  & 2  & 0  & 0\tabularnewline
0  & 0  & 0  & 4\tabularnewline
0  & 0  & 0  & 4\tabularnewline
\hline 
\end{tabular}J=1 %
\begin{tabular}{rrrr}
Energy  & T  & t  & v\tabularnewline
\hline 
0  & 1  & 0  & 2\tabularnewline
0  & 1  & 1  & 4\tabularnewline
0  & 1  & 1  & 4\tabularnewline
\hline 
\end{tabular}J=2 %
\begin{tabular}{rrrr}
Energy  & T  & t  & v\tabularnewline
\hline 
-1.25  & 0  & 1  & 2\tabularnewline
0  & 0  & 0  & 4\tabularnewline
0  & 0  & 0  & 4\tabularnewline
0  & 0  & 0  & 4\tabularnewline
-1.0  & 1  & 1  & 2\tabularnewline
0  & 1  & 1  & 4\tabularnewline
0  & 1  & 1  & 4\tabularnewline
-0.5  & 2  & 1  & 2\tabularnewline
0  & 2  & 2  & 4\tabularnewline
\hline 
\end{tabular}\\

\raggedright

\section{Results}

The Gamow-Teller operator is $C\sigma$t$_{+}$. The wave functions
for the Scandium isotopes are of the form \\
\begin{gather}
\sum\text{D}(J_{n}v){[}j_{p},J_{n}{]}^{I}
\end{gather}
with $j_{p}$, the angular momentum of the single proton equal to
7/2. Here D($J_{n}v$) is the probability amplitude that the neutrons
couple to $J_{n}$. The matrix element from McCullen {[}5{]}et al.
is 
\begin{gather}
M_{ij}=\sum\text{D}^{i}(j,J_{n})\text{D}^{f}(j,J_{n})U(1jJ_{f}J_{n};jJ_{i})
\end{gather}

We put the results of the calculated matrix elements in Table 4.1.

\subsection*{Table 4.1 {*}{*}{*}{*}}

\centering 
\begin{tabular}{rrrrr}
\hline 
7/2-7/2  & E(0)  & MBZE  & E(7)  & Q.Q\tabularnewline
\hline 
$^{43}$Sc  & 0.3849  & -0.2088  & -0.10160  & 0.1207\tabularnewline
$^{45}$Sc  & 0.2666  & 0.0927  & -0.0027  & 0.0255\tabularnewline
7/2-5/2(43)  & zero  & 0.2020  & 0.2902  & 0.2763\tabularnewline
7/2-5/2(45)  & zero  & 0.0459  & -0.0022  & 0.000792\tabularnewline
7/2-9/2(43)  & zero  & -0.0818  & 0.0168  & 0.008380\tabularnewline
7/2-9/2(45)  & zero  & 0.0008  & -0.0028  & -0.02399\tabularnewline
\hline 
\end{tabular}


\raggedright 

The results for the 7/2$^{+}$ to 7/2$^{-}$ transitions are shown
in the first two rows above. We see that J=0 T=1 pairing gives the
largest matrix element, MBZE is in the middle and E(J$_{\text{max}}$)
the smallest. Thus, we have the systematic that deviations for J=0
T=1 pairing lead to reduced Gamow-Teller matrix elements. It is not
surprising that the realistic case, MBZE, is in the middle because
the two-body interaction used in that calculation has both an a low
lying J=0 part but also a low lying J=7 part. Of perhaps greatest
interest is the fact that the matrix elements of GT for the E(0) interaction
vanish when $J_{f}$ is different than $J_{i}$. We have here considered
the cases $J_{i}$= (7/2)$_{1}$ and $J_{f}$= 5/2 or 9/2, both for
$^{43}$Sc and $^{45}$Sc . There is considerable discussion of the
pairing interaction in the 1993 book by Talmi{[}9{]}. He has a discussion
of odd tensor operators in space and spin. It is there shown that
these operators conserve seniority. In this work on GT we have a product
of an odd tensor operator in spin and an odd tensor operator in isospin.
The general selection rules for overall isospin are that T$_{f}$
can be equal to T$_{i}$, T$_{i}$ +1 or T$_{i}$ $-$1; although
in the cases considered here, the latter does not apply. We will soon
see that in general the GT operator does not conserve seniority. For
the J=0 T=1 pairing interaction the lowest state in $^{43}$Sc with
$J_{i}$=$j$=7/2 has seniority v=1. All other states for this and
all other angular momenta have v=3 except the T=3/2 J=$j$ state which
also has v=1. In the f$_{7/2}$ shell the latter state is unique.
We see from Table 4.1 that if our initial state is a v=1 state with
J=$j$ (7/2 in this case) and isospin T=1/2 there is a non-vanishing
matrix element to a v=1 T=3/2 state and J$_{f}$=$j$. However, with
a J=0 T=1 pairing interaction the matrix element from the v=1 state
to v=3 states with J $=j+1$ or J $=j-1$ vanish. It should be noted
that, although one constructs a $\text{J}=j$, v=1 state in say $^{43}$Sc
by first adding two neutrons coupled to $J_{n}=0$ to the single proton;
that is not the end of the story. One must introduce isospin wave
functions and antisymmetrize. The values of D$(J_{n}$) for the v=1
$\text{J}=j$ T=1/2 state for $J_{n}=0,2,4$ and 6 are respectively
(0.8660,0.2152,2887,0.3469). Consider the matrix element 
\begin{gather}
M'=N\left(\psi^{J_{f}}\sum\sigma t_{+}(1-P_{12}-P_{13})\left[j(1)\left[j(2)j(3)\right]^{0}\right]^{j}p(1)n(2)n(3)\right)
\end{gather}
where t$_{z}=-1/2$ for a proton and +1/2 for a neutron. We can replace
$\sum\sigma$t$_{+}$ by 3$\sigma$(1)t$_{+}$(1). Since t$_{+}$n=0
we see that the ($-P_{12}-P_{13}$) terms will not contribute. We
are left with $3N(j[\sigma j]^{j})(\psi^{J_{f}}{[}j(1){[}j(2)j(3){]}^{0}{]}^{j})$
and we can write $\psi^{J_{f}}$= $\sum\text{D}^{J_{f}}(J_{n}v){[}j_{p},J_{n}{]}^{J_{f}}$.
Hence the last factor is simply D$^{J_{f}}$(0). However, for a seniority
v=3 final state, D$^{J_{f}}$(0) is equal to zero. As mentioned before
the only T=3/2 state with seniority v=1 is the one with $J_{f}=j$.
The J=5/2 and 9/2 states all have v=3 and hence the matrix element
$M'$ vanishes for those cases, but there is a problem. The state
on the right is a mixture of J=7/2 v=1 T=1/2 and J=7/2 v=1 T=3/2.
We next show that the T=3/2 part also vanishes and this will imply
that the T=1/2 part will also vanish. Consider a transition from J=7/2$^{-}$
v=1 T=3/2 in $^{43}$Sc to J=5/2$^{-}$ or 9/2$^{-}$ with v=3 in
$^{43}$Ca. There is a close relation between Gamow-Teller transitions
and isovector magnetic dipole (M1) transitions. If one removes the
orbital part of the M1, keeping only the spin there is an isospin
relation between the two transitions. We can transform the GT problem
to one of M1 transitions in $^{43}$Ca. But it is well known that
for a single $j$ shell of particles of one kind, i.e. only neutrons,
all M1 transitions vanish.

We had previously displayed a formula for single $j$ shell M1 transitions
from an I=0$^{+}$ ground state to an I=1$^{+}$ state of an even-even
nucleus{[}9{]}.{*}{*}{*}{*}

This can be generalized to an expression given in the appendix (11).

Note that the term with J$_{p}$=0 does not contribute. From this
and the previous discussion on GT we see that it will also vanish
for a J=0 v=0 to J=1 v=4 {[}15{]}. {*}{*}{*}{*}

(Note that this expression implies that isoscalar transitions vanish
in the single $j$ shell limit i.e. $g_{p}-g_{n}=0$). In ref {[}2{]}
the energy shifts and B(GT)\textasciiacute s starting from the initial
J=0 $v$=0 T=0 state in $^{44}$Ti were given, although no proof of
the selection rule was given.

\section{Results in $^{44}$Ti}

We show the calculated B(M1) results here. Along the vertical we have
the I=1 states, and along the horizontal are the I=0 states.

\subsection*{Table 5.1 B(M1) Values in $^{44}$Ti for I=1 to I=0 Pairing Interaction}

\centering %
\begin{tabular}{rrrrr}
\hline 
State(v,T,t)  & $000$  & $400$  & $400$  & $020$\tabularnewline
\hline 
$210$  & 2.69963  & 8.0995  & 1.92994  & 0.898554\tabularnewline
$411$  & 0  & 7.6793  & 0.11174  & 0 \tabularnewline
$411$  & 0  & 1.91866  & 2.89221  & 0 \tabularnewline
\hline 
\end{tabular}\\

\raggedright

\subsection*{Table 5.2 B(M1) Values in $^{44}$Ti for I=1 to I=2 Pairing Interaction}

\centering %
\begin{tabular}{rrrrrrrrrr}
\hline 
State(v,T,t)  & $201$  & $400$  & $400$  & $400$  & $211$  & $411$  & $411$  & $221$  & $422$\tabularnewline
\hline 
$210$  & 1.02858  & 17.5613  & 0.0475777  & 2.29634  & 0  & 0  & 0  & 5.14334  & 0 \tabularnewline
$411$  & 0.181872  & 1.45084  & 0.0330685  & 1.8904  & 0  & 0  & 0  & 0.909075  & 8.2364 \tabularnewline
$411$  & 0.525607  & 1.4562  & 2.07128  & 3.32567  & 0  & 0  & 0  & 2.62751  & 0.465319 \tabularnewline
\hline 
\end{tabular}\\
 \newpage{}\raggedright

\subsection*{Table 5.3 B(M1) Values in $^{44}$Ti for I=1 to I=0 Q.Q Interaction}

\centering %
\begin{tabular}{rrrrrr}
\hline 
I  & $0_{1}$  & $0_{2}$  & $0_{3}$  & $0_{4}$  & \tabularnewline
\hline 
$1_{1}$  & 1.31736  & 1.80213  & 0.183327  & 0.0413715  & \tabularnewline
$1_{2}$  & 0.00146367  & 6.14543  & 9.041392  & 0.057738  & \tabularnewline
$1_{3}$  & 0.000661312  & 0.153487  & 0.953011  & 0.205204  & \tabularnewline
\hline 
\end{tabular}\\

\raggedright

\subsection*{Table 5.4 B(M1) Values in $^{44}$Ti for I=1 to I=2 Q.Q Interaction}

\centering %
\begin{tabular}{rrrrrrrrrr}
\hline 
I  & $2_{1}$  & $2_{2}$  & $2_{3}$  & $2_{4}$  & $2_{5}$  & $2_{6}$  & $2_{7}$  & $2_{8}$  & $2_{9}$\tabularnewline
\hline 
$1_{1}$  & 0.885292  & 0  & 5.0018  & 0.0301273  & 0.0533539  & 0  & 0.0781581  & 3.36014  & 0 \tabularnewline
$1_{2}$  & 0.0126882  & 0  & 3.30166  & 18.1444  & 8.08602  & 0  & 0.0339801  & 0.347936  & 0\tabularnewline
$1_{3}$  & 0.0000924735  & 0  & 0.180103  & 0.27692  & 0.534714  & 0  & 5.13135  & 8.26883  & 0 \tabularnewline
\hline 
\end{tabular}\\

\raggedright

We see that with the J=0 pairing interaction there is a nonzero transition
from a J=0 v=0 state to a J=1 v=2 state i.e. the M1 (or GT) operator
does not conserve seniority. We can, in analogy with what we did for
Scandium, form a $^{44}$Ti state ${[}{[}jj{]}^{0}{[}jj{]}^{0}{]}^{0}$
and antisymmetrize. This will be an admixture of J=0 v=0 T=0 and J=0
v=0 T=2. We now have to show that the T=2 part vanishes when we overlap
with a J=1 v=4 T=1 state and this will lead to the desired result
that the T=0 part vanishes. It is easier to use an isospin transformation
and consider the transition between a unique J=0 v=0 T=2 state in
$^{44}$Ca to a v=4 T=1 state in $^{44}$Sc. The T=2 state can be
obtained by forming the four neutron state ${[}{[}jj{]}^{0}{[}jj{]}^{0}{]}^{0}$
and antisymmetrizing. However, as shown before, we do not have to
antisymmetrize in the matrix element. And clearly; the v=4 T=1 J=1$^{+}$
state will, even after antisymmetrization not have any ${[}{[}jj{]}^{1}{[}jj{]}^{0}{]}^{1}$
component. Thus, the T=2 part vanishes and so will the T=0 part. We
will discuss the selection rules more systematically in the next section.

\section{More Results $-$ A Systematic Look at B(M1) Selection Rules for
$^{44}$Ti and $^{46}$Ti}

In the following tables, we make a systematic study of selection rules
for B(M1) transitions with a J=0 T=1 pairing interaction. The states
are classified by the quantum numbers (v,T,t) -seniority, isospin,
and reduced isospin. We have already presented the values of B(M1)
in ${^{44}}$Ti Tables 5.1 and 5.2 $-$ $^{44}$Ti I=1 to I=0 and
I=1 to I=2 pairing. Table 6.1 refers to large transitions in $^{44}$Ti
I=1 to 0, while Table 6.2 to its vanishing B(M1)'s. Tables 6.3 and
6.4 are for I=1 to I=2. Along the horizontal we have the I=1 states
and along the vertical the I=0 and I=2 states. Analogously, Tables
6.5 to 6.10 are presented for $^{46}$Ti.

\subsection*{Table 6.1 Large Values ($\geq$.5) for $^{44}$Ti I=1 to I=0}

{*}{*}{*}{*}remove? \centering $\begin{array}{rrr}
\hline \text{I=1} & \text{I=0} & \text{Value}\\
\hline 210 & 000 & 2.6693\\
210 & 020 & 8.0995\\
210 & 400 & 1.9299\\
210 & 400 & 0.8986
\\\hline \end{array}\begin{array}{rrr}
\hline \text{I=1} & \text{I=0} & \text{Value}\\
\hline 411 & 400 & 7.6793\\
411 & 400 & 2.8922\\
411 & 400 & 1.9187
\\\hline \end{array}$\\
 \raggedright

\subsection*{Table 6.2 Selection Rules $^{44}$Ti I=1 to I=0}

\centering $\begin{array}{rrr}
\hline \text{Selection Rule} & \text{I=1} & \text{I=0}\\
\hline \Delta v=4 & 411 & 000\\
\Delta v=4 & 411 & 020
\\\hline \end{array}$

\raggedright

\subsection*{Table 6.3 Large Values ($\geq$.5) for $^{44}$Ti I=1 to I=2}

\centering $\begin{array}{rrr}
\hline \text{I=1} & \text{I=2} & \text{Value}\\
\hline 210 & 201 & 1.02858\\
210 & 400 & 17.5613\\
210 & 400 & 2.29634\\
210 & 221 & 5.14334\\
411 & 400 & 1.45084\\
411 & 400 & 1.8904\\
411 & 221 & 0.909075
\\\hline \end{array}\begin{array}{rrr}
\hline \text{I=1} & \text{I=2} & \text{Value}\\
\hline 411 & 422 & 8.2364\\
411 & 201 & 0.525607\\
411 & 400 & 1.4562\\
411 & 400 & 2.07128\\
411 & 400 & 3.32567\\
411 & 221 & 2.62751
\\\hline \end{array}$ \\
 \raggedright

\subsection*{Table 6.4 Selection Rules $^{44}$Ti I=1 to I=2}

\centering $\begin{array}{rrr}
\hline \text{Selection Rule} & \text{I=1} & \text{I=0}\\
\hline \text{T}=1\Rightarrow\text{T}=1 & 210 & 211\\
\text{T}=1\Rightarrow\text{T}=1 & 210 & 411\\
\Delta\text{v}=2,\Delta\text{t}\neq0 & 210 & 422\\
\text{T}=1\Rightarrow\text{T}=1 & 411 & 211\\
\text{T}=1\Rightarrow\text{T}=1 & 411 & 211
\\\hline \end{array}$\\

\raggedright

\subsection*{Table 6.5 $^{46}$Ti I=1 to I=0}

\centering $\begin{array}{cccccccc}
\hline \text{State(v,T,t)} & 411 & 411 & 611 & 611 & 220 & 421 & 421\\
\hline 010 & 0 & 0 & 0. & 0. & 1.0799 & 0 & 0\\
410 & 2.8794 & 0.0491 & 0. & 0. & 2.4344 & 0.5611 & 0.4150\\
410 & 0.7573 & 5.7648 & 0. & 0. & 0.3947 & 0.1157 & 2.0588\\
611 & 1.0423 & 0.0987 & 2.3989 & 0.6317 & 0. & 3.1539 & 0.2640\\
611 & 0.0049 & 0.1721 & 0.0001 & 1.7267 & 0. & 0.0858 & 0.4450\\
030 & 0 & 0 & 0. & 0. & 9.7201 & 0 & 0
\\\hline \end{array}$\\
 \raggedright

\subsection*{Table 6.6 Large Values ($\geq$.5) for $^{46}$Ti I=1 to I=0}

\centering $\begin{array}{rrr}
\hline \text{I=1} & \text{I=0} & \text{Value}\\
\hline 411 & 410 & 2.8794\\
411 & 410 & 0.7573\\
411 & 611 & 1.0423\\
411 & 410 & 5.7648\\
611 & 611 & 2.3989\\
611 & 611 & 0.6317\\
611 & 611 & 1.7267
\\\hline \end{array}\begin{array}{rrr}
\hline \text{I=1} & \text{I=0} & \text{Value}\\
\hline 220 & 010 & 1.0799\\
220 & 410 & 2.4344\\
220 & 030 & 9.7201\\
421 & 410 & 0.5611\\
421 & 611 & 3.1539\\
421 & 410 & 2.0588
\\\hline \end{array}$\\
 \raggedright

\subsection*{Table 6.7 Selection Rules $^{46}$Ti I=1 to I=0}

\centering $\begin{array}{rrr}
\hline \text{Selection Rule} & \text{I=1} & \text{I=0}\\
\hline \Delta\text{T}=2,\hspace{0.2cm}\Delta\text{v}=4 & 411 & 030\\
\Delta\text{T}=2,\hspace{0.2cm}\Delta\text{v}=4 & 611 & 030\\
\Delta\text{v}=4 & 411 & 010\\
\Delta\text{v}=4 & 421 & 010\\
\Delta\text{v}=6 & 611 & 010\\
\Delta\text{v}=4 & 220 & 611\\
\Delta\text{v}=4 & 421 & 030?\\
\Delta\text{v}=2,\Delta\text{t}\neq0 & 611 & 410
\\\hline \end{array}$ \newpage{}\raggedright

\subsection*{Table 6.8 $^{46}$Ti I=1 to I=2}

\centering $\begin{array}{cccccccc}
\hline \text{State(v,T,t)} & 411 & 411 & 611 & 611 & 220 & 421 & 421\\
\hline 211 & 0.9874 & 0.3326 & 0 & 0 & 1.3712 & 0.0272 & 0.0019\\
211 & 0.4367 & 0.1472 & 0 & 0 & 0.1715 & 0 & 0.3238\\
412 & 0.0916 & 1.5360 & 0. & 0. & 0 & 0.0607 & 0.4819\\
411 & 0.0847 & 0.0914 & 0.4365 & 0.0065 & 0. & 0.0374 & 0.0261\\
411 & 0.0041 & 0.0186 & 1.5191 & 0.0152 & 0. & 0.0846 & 0.0668\\
410 & 0.0646 & 1.6850 & 0. & 0. & 12.1303 & 0.0832 & 0.5004\\
410 & 3.5617 & 0.1189 & 0. & 0. & 2.9785 & 0.6431 & 0.5838\\
410 & 0.4668 & 2.4445 & 0. & 0. & 5.3986 & 0.0273 & 0.9432\\
611 & 2.1377 & 0.2523 & 2.3618 & 0.0555 & 0. & 2.9801 & 0.4370\\
611 & 0.2654 & 0.0135 & 0.1597 & 0.8390 & 0. & 0.0329 & 0.2333\\
611 & 0.0616 & 0.1344 & 7.1099 & 1.4178 & 0. & 1.4482 & 0.5082\\
611 & 0.0375 & 0.0024 & 0.0873 & 0.0461 & 0. & 0.0123 & 0.0127\\
611 & 0.1215 & 1.3291 & 0.0001 & 5.7321 & 0. & 0.0315 & 4.0036\\
221 & 2.2323 & 0.7524 & 0 & 0 & 2.5716 & 0.0398 & 0.0883\\
422 & 0.2746 & 4.6069 & 0. & 0. & 0 & 0.1821 & 1.4454\\
421 & 0.1804 & 0.0338 & 0.6123 & 0.0630 & 0. & 0.3563 & 0.0188\\
421 & 0.0862 & 0.2962 & 5.2534 & 0.0019 & 0. & 1.2615 & 0.2597\\
231 & 0 & 0 & 0 & 0 & 2.0572 & 0.5125 & 4.5230
\\\hline \end{array}$ \\
 \raggedright

\subsection*{Table 6.9 Large Values ($\geq$.5) for $^{46}$Ti I=1 to I=2}

\centering $\begin{array}{rrr}
\hline \text{I=1} & \text{I=2} & \text{Value}\\
\hline 411 & 211 & 0.9873\\
411 & 410 & 3.5617\\
411 & 611 & 2.1377\\
411 & 221 & 2.2323\\
411 & 412 & 1.5360\\
411 & 410 & 1.6850\\
411 & 410 & 2.4444\\
411 & 611 & 1.3291\\
411 & 221 & 0.7524
\\\hline \end{array}\begin{array}{rrr}
\hline \text{I=1} & \text{I=2} & \text{Value}\\
\hline 411 & 422 & 4.6069\\
611 & 411 & 1.5190\\
611 & 611 & 2.3617\\
611 & 611 & 7.1099\\
611 & 421 & 0.6122\\
611 & 421 & 5.2534\\
611 & 611 & 1.4178\\
611 & 611 & 5.7321
\\\hline \end{array}\begin{array}{rrr}
\hline \text{I=1} & \text{I=2} & \text{Value}\\
\hline 220 & 211 & 1.3712\\
220 & 410 & 12.1303\\
220 & 410 & 2.9784\\
220 & 410 & 5.3986\\
220 & 221 & 2.5716\\
220 & 231 & 2.0572\\
421 & 410 & 0.6431\\
421 & 611 & 2.9801\\
421 & 611 & 1.4482
\\\hline \end{array}\begin{array}{rrr}
\hline \text{I=1} & \text{I=2} & \text{Value}\\
\hline 421 & 421 & 1.2615\\
421 & 231 & 0.5125\\
421 & 410 & 0.5838\\
421 & 410 & 0.9432\\
421 & 611 & 0.5082\\
421 & 611 & 4.0036\\
421 & 422 & 1.4454\\
421 & 231 & 4.5230
\\\hline \end{array}$\\

\raggedright

\subsection*{Table 6.10 Selection Rules $^{46}$Ti I=1 to I=2}

\centering $\begin{array}{rrr}
\hline \text{Selection Rule} & \text{I=1} & \text{I=0}\\
\hline \Delta\text{T}=2 & 411 & 231\\
\Delta\text{T}=2,\hspace{0.2cm}\Delta\text{v}=4 & 611 & 231\\
\Delta\text{v}=4 & 611 & 211\\
\Delta\text{v}=4 & 220 & 611\\
\Delta\text{v}=2,\Delta\text{t}\neq0 & 611 & 412\\
\Delta\text{v}=2,\Delta\text{t}\neq0 & 611 & 410\\
\Delta\text{v}=2,\Delta\text{t}\neq0 & 611 & 422\\
\Delta\text{v}=2,\Delta\text{t}\neq0 & 220 & 412\\
\Delta\text{v}=2,\Delta\text{t}\neq0 & 220 & 411\\
\Delta\text{v}=2,\Delta\text{t}\neq0 & 220 & 422\\
\Delta\text{v}=2,\Delta\text{t}\neq0 & 220 & 421\\
\\
\hline \end{array}$\\
 \raggedright

\section{Discussion of the Tables}

We observe that B(M1)'s vanish in the following cases:

a.) In $^{44}$Ti B(M1)'s (N=Z) from T=1 to T=1 vanish.

b.) $\Delta$T=2 or more

c.) $\Delta$v=4 or 6

d.) $\Delta\text{v}=2$ and $\Delta\text{t}\neq0$

The selection rule for case a.) is well known. It is discussed in
several places including the book by Talmi {[}9{]}. It can be explained
by the vanishing of the Clebsh-Gordan coefficient (1,1,0,0|1,0).

For case b.) where the change of isospin is 2 or more units is also
an easy to explain. These B(M1)'s obviously are zero because the M1
operator is of rank 1 in isospin. Some examples are: $(411)^{2}\rightarrow(030)$,
$(611)^{2}\rightarrow(030)$.

In case c.) the change of seniority is more that 2 units i.e 4 or
6. The B(M1)'s for these cases also obviously are zero because the
one body M1 operator can only uncouple one J=0 pair. Some examples
are: $(411)^{2}\rightarrow(010)$ , $(421)^{2}\rightarrow(010)$,
$(611)^{2}\rightarrow(010)$, $(220)^{2}\rightarrow(611)$, $(421)^{2}\rightarrow(030)$.

In case d.) we get vanishing B(M1)'s when seniority and the reduced
isospin simultaneously change. The M1 operator can attack a J=0 pair
and increase the isospin but that will not affect the the particles
not coupled to zero whose isospin is indeed the reduced isospin. Examples
of this are in the Sc isotopes where J=7/2 T=1/2 transitions to J=5/2
or 9/2 states with T=3/2 are forbidden with a pairing interaction.
Also in the Ti isotopes (611)$^{2}\rightarrow(412),(410)^{3},(422)$,
and $(220)\rightarrow(412),(411)^{2},(422),(421)^{2}$.

There is one ambiguity-the case I=1$^{+}$ to 2$^{+}$; there are
two (421) states. One has a non-zero B(M1) to (211) and the other
does not. However, when there is a 2-fold degeneracy one can take
arbitrary linear combinations of the two states and so get a variety
of B(M1)'s. 

We mention briefly that Zamick {[}11{]} had previously considered
M1 transitions from J=0$^{+}$ ground states in Ti isotopes to J=1$^{+}$
excited states in the context of scissor modes. These transitions
are sometimes called spin scissors excitations and have a fair amount
of orbital content-not just spin. They bear some analogy to the scissors
modes in deformed nuclei such as $^{156}$Gd {[}12{]}.

In the appendix we give detailed expressions for B(M1)'s and B(GT).
It should be noted that such a relation between them has been previously
discussed by L. Zamick and D.C. Zheng {[}13{]}, but not in such a
complete way.

Matthew Harper thanks the Rutgers Aresty Research Center for Undergraduates
for support during the 2014-2015 fall-spring session.We thank Kai
Neergaard for helpful comments.

\section{Appendix}

\subsection*{Formulas For B(GT)}

\begin{gather}
X_{1}=\sum\limits _{J_{p}J_{n}}D^{f}(J_{p}J_{n})D^{i}(J_{p}J_{n})U(1J_{p}I_{f}J_{n};J_{p}I_{i})\sqrt{J_{p}(J_{p}+1)}\\
X_{2}=\sum\limits _{J_{p}J_{n}}D^{f}(J_{p}J_{n})D^{i}(J_{p}J_{n})U(1J_{n}I_{f}J_{p};J_{n}I)\sqrt{J_{n}(J_{n}+1)}\\
B(GT)=0.5\frac{2I_{f}+1}{2I_{i}+1}f(j)^{2}\left[\frac{\langle1T_{i}1M_{T_{i}}|T_{f}M_{T_{f}}\rangle}{\langle1T_{i}0M_{T_{i}}|T_{f}M_{T_{i}}\rangle}\right]^{2}(X_{1}-(-1)^{I_{f}-I_{i}}X_{2})^{2}\\
\text{Where }f(j)=\begin{cases}
\frac{1}{j} & \text{if }j=l+1/2\text{ e.g }f_{7/2}\\
\frac{-1}{j+1} & \text{if }j=l-1/2\text{ e.g }f_{5/2}
\end{cases}\\
ft=\frac{6177}{B(F)+1.583B(GT)}
\end{gather}

\subsection*{Formulas For B(M1)}

\begin{gather}
B(M1)=\frac{3}{4\pi}\frac{2I_{f}+1}{2I_{i}+1}\left[g_{j_{p}}X_{1}+(-1)^{I_{f}-I_{i}}g_{j_{n}}X_{2}\right]^{2}\\
\text{Here }g_{j}=g_{l}\pm\left\lbrace \frac{g_{s}-g_{l}}{2l+1}\right\rbrace \\
\begin{align}g_{s_{p}}=5.586 &  & g_{l_{p}}=1\\
g_{s_{n}}=-3.826 &  & g_{s\l_{n}}=0
\end{align}
\end{gather}

For the case T$_{f}$ is not equal to T$_{i}$ we find: 
\begin{gather}
X_{1}=(-1)^{I_{f}-I_{i}+1}X_{2}\\
B(M1)=\frac{3}{4\pi}\frac{2I_{f}+1}{2I_{i}+1}(g_{jp}-g_{jn})^{2}X_{1}^{2}\\
B(GT)=2\frac{2I_{f}+1}{2I_{i}+1}f(j)^{2}\left[\frac{\langle1T_{i}1M_{T_{i}}|T_{f}M_{T_{f}}\rangle}{\langle1T_{i}0M_{T_{i}}|T_{f}M_{T_{i}}\rangle}\right]^{2}(X_{1})^{2}
\end{gather}

With this simplification we see that B(GT) is proportional to B(M1).

Using bare values we find B(GT)/B(M1)= 0.1411 for j= 7/2 in $^{44}$Ti.

The magnetic moment is: 
\begin{gather}
\frac{\mu}{I}=\frac{g_{j_{p}}+g_{j_{n}}}{2}+\frac{g_{j_{p}}-g_{j_{n}}}{2(I+1)}\left[\sum_{J_{p}J_{n}}|D(J_{p}J_{n})|^{2}\left[J_{p}(J_{p}+1)-J_{n}(J_{n}+1)\right]\right]
\end{gather}

\end{document}